\documentclass{aa}  
\usepackage{graphicx}
\usepackage{txfonts}
\usepackage{hyperref}
\usepackage{subcaption}
\usepackage{float}

\usepackage{multirow}
\usepackage{natbib}
\bibpunct{(}{)}{;}{a}{}{,} 
\usepackage{orcidlink}

\usepackage{soul} 

\begin{document}

\title{Exploring Galactic open clusters with Gaia}
\subtitle{I. An examination in the first kiloparsec}


\author{Jeison Alfonso\inst{1} \orcidlink{0009-0004-5488-4945}
      \and
      Alejandro Garc\'{i}a-Varela\inst{1} \orcidlink{0000-0001-8351-0628}
      \and 
      Katherine Vieira\inst{2} \orcidlink{0000-0001-5598-8720}
      }

\institute{Universidad de los Andes,
          Departamento de F\'{i}sica,
          Cra. 1 No. 18A-10, Bloque Ip, A.A. 4976
          Bogot\'{a}, Colombia \\
          \email{je.alfonso1@uniandes.edu.co}
          \and
          Instituto de Astronom\'{i}a y Ciencias Planetarias, 
          Universidad de Atacama, Copayapu 485, 
          Copiap\'{o} 1531772, Chile
          }
          
\date{Received date / Accepted date}

  \abstract
   {Since the first publication of the {\it Gaia} catalogue a new view of our Galaxy has arrived. Its astrometric and photometric information has improved the precision of the physical parameters of open star clusters obtained from them.}
   {Using the {\it Gaia} DR3 catalogue, we aim to find physical stellar members including faint stars for $370$ Galactic open clusters located within $1$ kpc. We also estimate the age, metallicity, distance modulus and extinction of these clusters.}
   {We employ the \texttt{HDBSCAN} algorithm on both astrometric and photometric data to identify members in the open clusters. Subsequently, we refine the samples by eliminating outliers through the application of the Mahalanobis metric utilizing the $\chi^2$ distribution at a confidence level of 95\%. Furthermore, we characterize the stellar parameters with the PARSEC isochrones.}
   {We obtain reliable star members for $370$ open clusters with an average parallax error of $\sigma_{\varpi} = 0.16$ mas. We identify about $\sim 40\%$ more stars in these clusters compared to previous work using the {\it Gaia} DR2 catalogue, including faint stars as new members with $G \geq 17$. Before the clustering application we correct the parallax zero-point bias to avoid spatial distribution stretching that may affect clustering results. Our membership lists include merging stars identified by \texttt{HDBSCAN} with astrometry and photometry. We note that the use of photometry in clustering can recover up to $10\%$ more stars in the fainter limit than clustering based on astrometry only, this combined with the selection of stars filtering them out by quality cuts significantly reduces the number of stars with huge $\sigma_{\varpi}$. After clustering, we estimate age, $Z$, and $A_V$ from the photometry of the membership lists.}
   {We have carried out a search to extent the membership list for $370$ open clusters mainly on the Galactic plane in a neighborhood of $1$ kpc. Our methodology provides a robust estimator for outliers identification and also extends the membership lists to fainter stars in most of the clusters. Our findings suggest the need to carefully identify spurious sources that may affect clustering results.}


\maketitle 


\section{Introduction}

The Milky Way disk has been the birthplace of open clusters that continually evolve over time \citep{lada2003,mckee2007}. 
The interaction between these stellar systems and the Galactic gravitational potential, shapes their collective dynamics allowing us to understand the evolutionary state of the Galaxy. 
As clusters age, their members gradually disperse into the surrounding increasing the field population. 
This phenomenon presents an opportunity to test stellar and gravitational models \citep{kupper2015}. 
The loss of stellar members takes place through the process of relaxation, sometimes also known as dissolution or evaporation \citep{krumholz2019}, resulting in the formation of elongated structures characterized by over and under-densities attributed to gravitational interactions with giant molecular clouds, or with the spiral arms or the bulge of the Milky Way \citep{binney2008}.
These stars leave the cluster through the Lagrangian points at velocities that slightly exceed the escape velocity \citep{kupper2008}.

Currently, the open cluster census has been increasing, largely attributed to the wealth of data provided by the {\it Gaia} Data Release 3 (DR3)  \citep{gaia2016,gaiadr32022,babusiaux2022}. According to \citet{hunt2023}, approximately seven thousand clusters have been identified. 
To determine the star members of these systems, one approach can be the application of unsupervised machine learning techniques based on density hierarchical algorithms, as done by \citet{castro-ginard2018}, \citet{cantat-gaudin2020} and \citet{castro-ginard2022} for Galactic open cluster.

\begin{figure*}[h]
    \centering
    \includegraphics[scale=0.70]{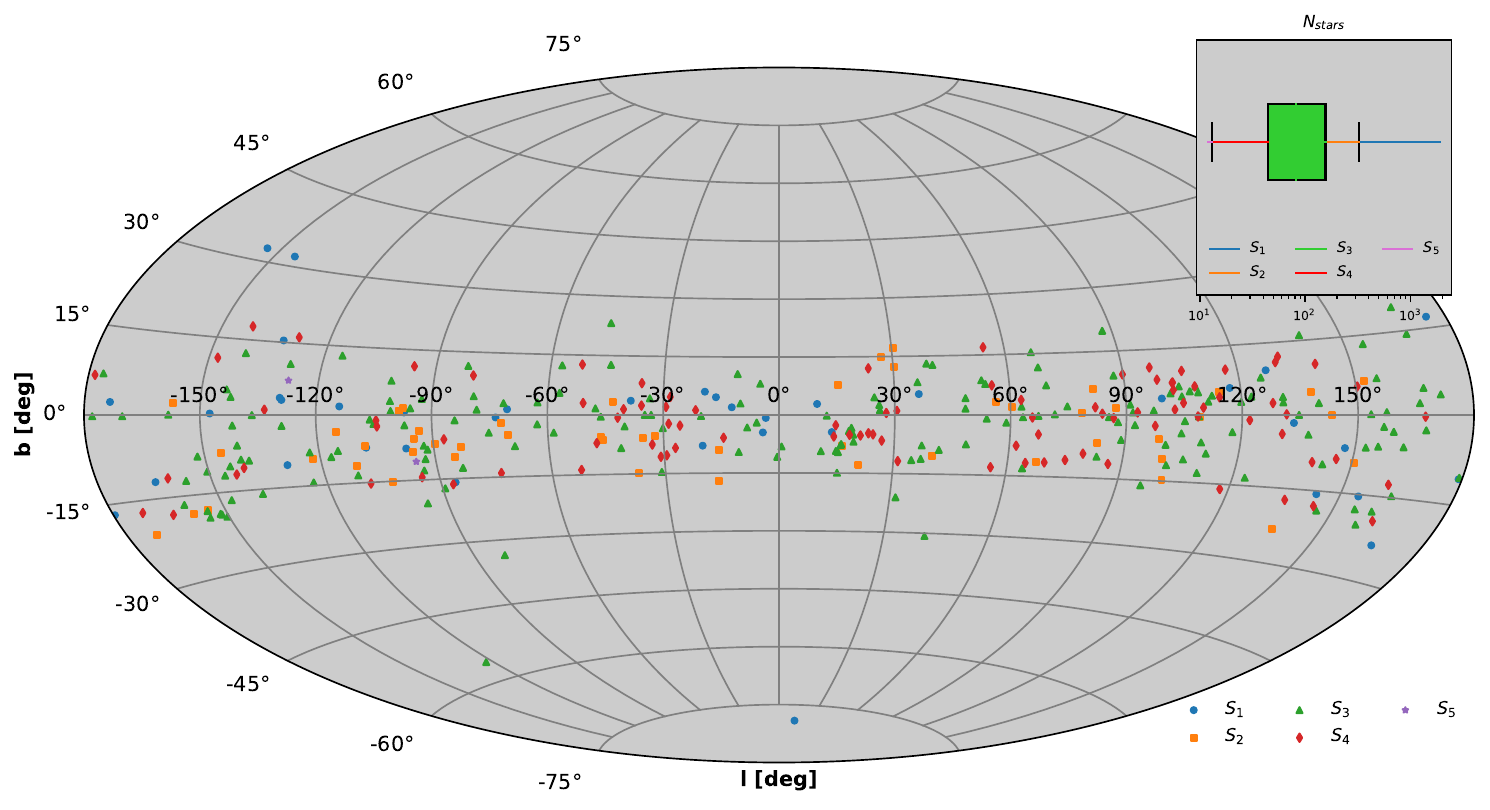}
    \caption{Celestial map in Galactic coordinates for $370$ open clusters selected from the \citet{cantat-gaudin2020} sample within a $1$ kpc neighborhood of the Sun. The colors represent open clusters belonging to some of the five sub-samples ($S_i$, for $i=1,\dots, 5$) defined from the box plot of their number of stars ($N_{\text{stars}}$). The limits between sub-samples correspond to $N_{\text{stars}}=324, 156, 44, 13$, respectively.}
    \label{fig:data_samples}
\end{figure*}

Age and metallicity play a fundamental role in understanding processes like mass segregation and expansion of open clusters \citep{croce2023}. 
Recent studies based on spectroscopic and photometric data have found different distributions of these parameters \citep{yong2012,anders2017,spina2017}. 
Once cluster members have been identified using Gaia data, their age and metallicity can be determined by isochrone fitting.

In this work we aim to obtain reliable stellar members and cluster parameters for $370$ nearby open clusters, constrained within $1$ kpc, using \cite{cantat-gaudin2020} as a reference catalogue. 
Before the clustering implementation, we correct the zero-point parallax bias for all sources based on the functions presented by \citet{lindegren2021}. 
Then, we utilize the \texttt{HDBSCAN} algorithm, computing distances with the Mahalanobis metric in two feature spaces: one comprised of equatorial coordinates, proper motions and parallaxes, and the other with two additional photometric colors compared to the previous set of features. 
The cluster members were selected as the union of the members found in each feature space. 
Next, we remove outliers using once again the Mahalanobis metric considering a $95\%$ confidence interval with the $\chi^2$ distribution. 
Finally, we estimate the age, metallicity, extinction, and distance modulus to each cluster fitting the PARSEC tracks \citep{bressan2012,marigo2013} to the Color-Magnitude Diagrams (CMDs) of the stellar members.

This document is structured in six sections, which are summarized below: 
Section \ref{sec:data} details the data selection within each open cluster region, based on equatorial coordinates and the criteria to ensure the inclusion of high-quality astrometric sources.
Section \ref{sec:methods}, comprises three subsections: a detailed explanation of the \texttt{HDBSCAN} algorithm, the methodology for determining membership, and the procedures for identifying and removing outliers.
Section \ref{sec:results} presents the outcomes of our analysis, including the determination of stellar membership within each cluster, challenges encountered in dealing with parallax uncertainties for faint stars, and the methodology employed for inferring cluster parameters.
Section \ref{sec:discussion} addresses the implications and significance of our results. Finally, our key conclusions and insights are summarized in Section \ref{sec:conclusions}.


\section{Data}
\label{sec:data}

The identification of star members in open clusters using {\it Gaia} data, typically involves employing a magnitude cutoff around $G \sim 17$ or $G \sim 18$ to mitigate the impact of increased astrometric errors, particularly pronounced for distant clusters \citep{castro-ginard2018, cantat-gaudin2020}. 
Nevertheless, with the {\it Gaia} DR3 data, it is now feasible to detect star members in open clusters even in the faintest regions of the CMD, while still maintaining reliable astrometric measurements.

Among the $2017$ open clusters in \citet{cantat-gaudin2020}, we select $370$ clusters within $1$ kpc. 
The Hyades and Coma clusters were excluded from our analysis due to their large apparent size caused by their proximity to the Sun. 
As the distribution of tidal radii of open clusters reported in the literature does not exceed $50$ pc, we adopt this value as the maximum physical size of any Galactic open cluster.
Taking into account the reported distance and coordinates of each cluster given by \cite{cantat-gaudin2020}, we computed the maximum angular size covered in the sky for each of these systems. 
We take this apparent size as the search radius in the {\it Gaia} query, centered on their equatorial coordinates to download data.
We also made a distance cutoff, centered on the mean cluster parallax, of $\pm 400$ pc in each region by converting it into parallax (mas) to avoid including field stars too distant from the cluster.
In addition, to remove spurious sources and ensure high-quality data, we filter out data requiring:

\begin{enumerate}
    \item \verb|ruwe| $< 1.4$,
    \item \verb|parallax_over_error| $>10$,
    \item \verb|visibility_periods_used| $>6$,
    \item \verb|astrometric_excess_noise| $<1$,
    \item \verb|phot_{g,bp,rp}_mean_flux_over_error| $>10$.
\end{enumerate}

Making the box plot of the number of members ($N_{\text{stars}}$) in each cluster reported by \citet{cantat-gaudin2020}, our sample of $370$ open clusters has been categorized into five distinct sub-samples. Sub-sample $S_1$ has the highest star counts clusters (38 clusters, $N_{\text{stars}} > 324$) while $S_5$ has the lowest ones ($2$ clusters, $N_{\text{stars}} \leq 13$).
Sub-sample $S_2$ comprises $53$ clusters with $156 < N_{\text{stars}} \leq 324$, while $S_4$ has $95$ clusters with $13 < N_{\text{stars}} \leq 44$. 
The central interquartile sub-sample is $S_3$ with $182$ clusters having $44 < N_{\text{stars}} \leq 156$. All of these sub-samples and the box plot are shown in Fig. \ref{fig:data_samples}.



\section{Methods}
\label{sec:methods}

In this section, we provide an overview of the clustering algorithm used to perform membership and the methodology to obtain star members in each open cluster, remove outliers, and infer their cluster parameters.


\subsection{\texttt{HDBSCAN}}
\label{sec:hdbscan}

The Hierarchical Density-Based Spatial Clustering of Applications with Noise \texttt{HDBSCAN} \citep{mcinnes2017} is suitable for application to tabular {\it Gaia} data. 
This algorithm is based on implementing \texttt{DBSCAN} \cite{ester1996} varying the epsilon value. 
\texttt{HDBSCAN} builds a hierarchy based on the minimum spanning tree graph from a mutual reachability distance computed with an input pairwise metric.
Then, it begins to remove edges from the dense graph and look for cluster stability: the most stable ones are classified as clusters \citep[and reference therein]{campello2013}.
Recently, \cite{hunt2021} showed that this algorithm has the best performance on {\it Gaia} data in reducing false positives compared to DBSCAN and Gaussian Mixture models, it can recover clusters with many shapes and sizes which allows to explore the substructures and halo populations in many clusters \citep{zhong2019}.
\texttt{HDBSCAN} has also significantly enriched the census of Galactic open clusters, enabling the discovery of new clusters and recovering the established ones in the literature.
The main hyperparameters in \texttt{HDBSCAN} are the minimum number of samples in a cluster (\verb+min_cluster_size+) and the number of samples in a neighborhood (\verb+min_samples+).
These hyperparameters need to be carefully selected to have the best efficiency running on {\it Gaia} data.

To obtain clusters from the data, the \texttt{HDBSCAN} algorithm requires a metric to compute distances between points in the dataset.
The Euclidean metric is the most commonly used, however, it requires that the correlation matrix has unit-variance \citep{feigelson2012}. 
This is true for independent data, but real-world data has different kinds of direct correlations.
Indeed, in the {\it Gaia} catalogue there are correlations between the astrometric data at different levels. 
As mentioned in \citet[see Sect. 4.5.7]{gaiadr32022}, correlations exist between astrometric parameters for the same source, between different sources for the same astrometric parameter, and between arbitrary astrometric parameters for different sources.
Therefore, these issues may affect the results of clustering algorithms.
One approach to deal with correlations between data is using the Mahalanobis metric to compute distances in the dataset \citep{mahalanobis}. 
To our knowledge, this metric has not been used with {\it Gaia} data. 
It is useful in multivariate analysis and can also be employed to identify outliers (see Sect. \ref{sec:outlier_removal}).

Suppose we have any two vectors $\Vec{r_i}$ and $\Vec{r_j}$ in the dimensional space with a positive-definite covariance matrix $\Vec{S}$, thus, the Mahalanobis distance $d_{M}(\Vec{r_i}, \Vec{r_j})$ between $\Vec{r_i}$ and $\Vec{r_j}$ is

\begin{equation}\label{equ:maha-metric}
    d_{M}(\Vec{r_i}, \Vec{r_j}) = \sqrt{(\Vec{r_i} - \Vec{r_j})^{T} \Vec{S}^{-1} (\Vec{r_i} - \Vec{r_j})}.
\end{equation}

This metric is passed to \texttt{HDBSCAN} as the pairwise metric, it takes into account any level of correlation in data and is closely to Hotelling's $T^2$ based on multivariate normal distribution \citep{feigelson2012}. 
It is worth mentioning that if the covariance matrix has unit-variance, e.i., the covariance matrix is the unity, it recovers to the well known standard Euclidean metric.


\subsection{Membership determination}
\label{sec:membership_determination}

The current membership lists of Galactic open clusters include few or no faint stars due to catalogue limitations. 
Recent works (e.g., \citealt{groeningen2023,hunt2023}) provide updated membership lists reaching the {\it Gaia} limit up to $G \sim 21$. 
At this level, the low-mass stars are strongly scattered from the main sequence, and this may affect cluster parameter estimates showing a clear need to be cautious at fainter magnitudes. 
Nevertheless, once we deal with error and correlation problems, it is thus possible to extend the star membership lists with reliable statistics even in the faint region of the CMDs.

\begin{figure*}[h]
    \centering
    \includegraphics[scale=0.8]{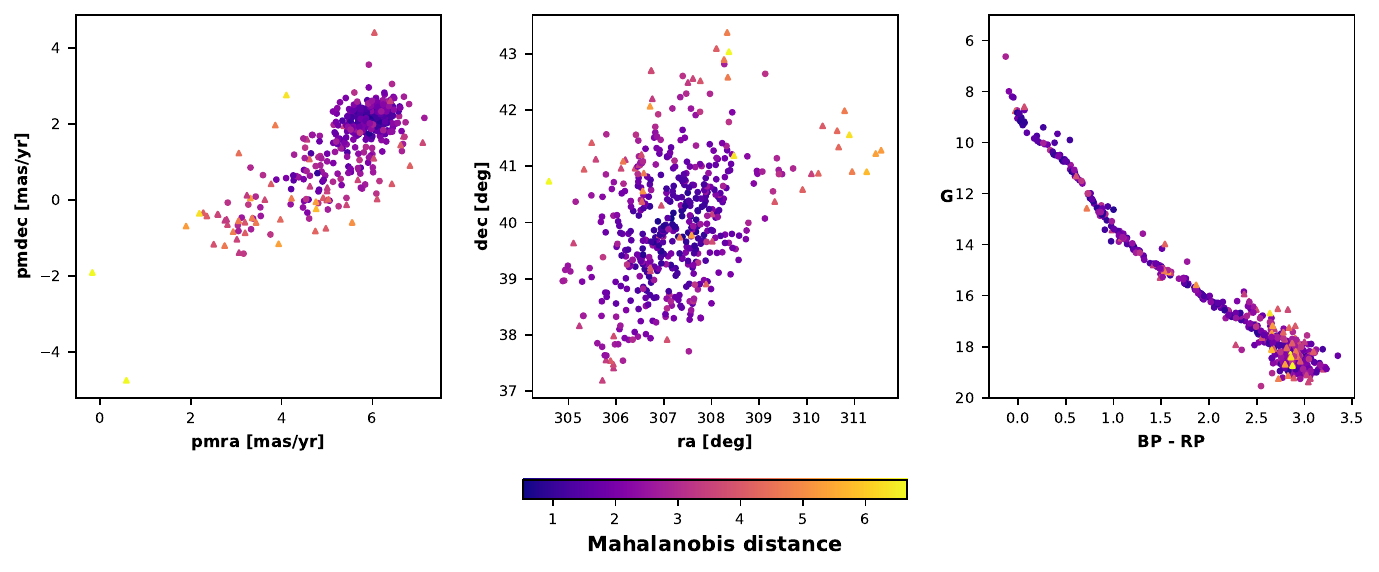}
    \caption{Vector Point Diagram (left panel), equatorial coordinates (middle panel) and CMD (right panel) of the 486 members selected by \texttt{HDBSCAN} for Roslund 6 open cluster. After an outlier threshold corresponding to $3.33$ following the methodology presented in Sect. \ref{sec:outlier_removal}, the sample reduces to $429$ with $57$ stars identified as outliers. 
    The cluster members and outliers are marked by circles and triangles, respectively.
    The color bar indicates the Mahalanobis distances.}
    \label{fig:cluster_outliers}
\end{figure*}

Before the clustering implementation, we correct the parallax zero-point bias given the ecliptic latitude, magnitude and colour of any source throughout the Python library \verb+gaiadr3_zeropoint+\footnote{\url{https://gitlab.com/icc-ub/public/gaiadr3_zeropoint}}, this allows us to obtain a corrected parallax ($\varpi_{P}$) for each star based on functions presented by \cite{lindegren2021}.

To recover the stellar members in each open cluster we adopt two different approaches to apply \texttt{HDBSCAN} using the Mahalanobis metric given by Eq. \eqref{equ:maha-metric}. 
First, we use the common astrometric space with five features, we name it set \verb+5F+: ($\alpha$, $\delta$, $\mu_{\alpha *}$\footnote{\label{mu_alpha}$\mu_{\alpha *} = \mu_{\alpha} \cos \delta$}, $\mu_{\delta}$, $\varpi_P$), then we use a combination of astrometry and photometry with seven features, \verb+7F+: ($\alpha$, $\delta$, $\mu_{\alpha *}$, $\mu_{\delta}$, $\varpi_P$, $G - G_{RP}$, $G_{BP} - G_{RP}$), where we include two color indexes. 
We note that adding photometry to perform clustering increases the number of fainter stars up to $G \sim 19.8$ in most of the clusters, but decreases the number of the brightest ones. 
Once the clustering algorithm has been performed on \verb+5F+ and \verb+7F+, we merge star members from the two sets (\verb+5F+$\cup$\verb+7F+) in order to have the larger amount of them.
This produces a final list that includes reliable stellar members up to $G \sim 19.8$ that were undetected by the astrometric space \verb+5F+ alone.


\subsection{Outlier removal}
\label{sec:outlier_removal}

Despite the quality cuts applied on {\it Gaia} DR3 data explained in Sect. \ref{sec:data}, and that \texttt{HDBSCAN} is very efficient and robust in detecting clusters with different shapes, the membership candidates are not exempt from selecting outliers \citep{feigelson2012,hunt2021}.
They can be bona fide extreme objects in a single population, the result of systematic errors or contamination by spurious data. 
To remove these problematic data, we also employ the Mahalanobis distance.
This metric cannot only be used as a pairwise metric as done in Sect. \ref{sec:hdbscan}, but also to remove outliers.
In this case, the Mahalanobis metric can be used as a measure of the distance between a point $\Vec{r_i}$ in the dataset and a probability distribution $Q$.
Following Eq. \eqref{equ:maha-metric}, for a vector $\Vec{r_i}$ the Mahalanobis distance is

\begin{figure}[h]
    \centering
    \includegraphics[scale=0.73]{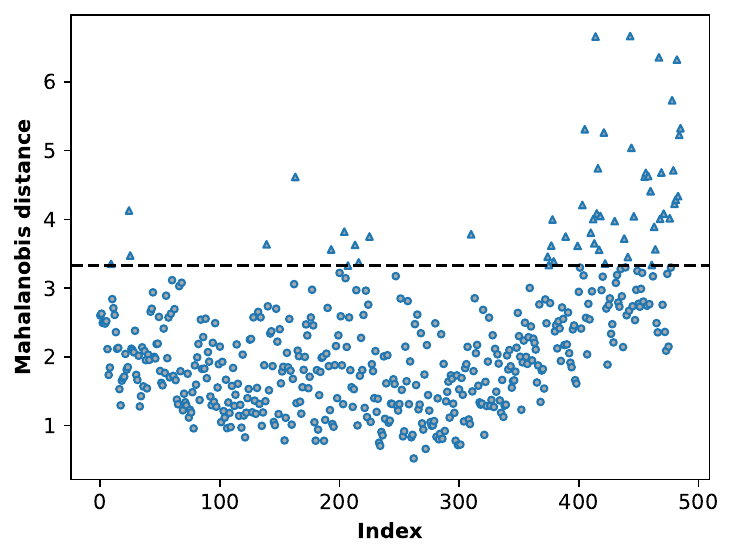}
    \caption{The Mahalanobis distance distribution for the Roslund 6 members selected by the \texttt{HDBSCAN} algorithm. The horizontal dashed black line is the outlier threshold with a value of $c = 3.33$. There are $57$ stars identified as outliers in this open cluster, marked by triangles above the dashed line.
    The horizontal axis represents an arbitrary star index assigned by \texttt{HDBSCAN} to the selected members.}
    \label{fig:roslund_6_maha}
\end{figure}

\begin{equation}\label{equ:maha-distance}
    d_{M}(\Vec{r_i}) = \sqrt{(\Vec{r_i} - \Vec{\mu})^{T} \Vec{S}^{-1} (\Vec{r_i} - \Vec{\mu})},
\end{equation}
where $\Vec{\mu}$ is the sample mean and $\Vec{S}$ is the positive-definite covariance matrix. 
In other words, the distance given by Eq. \eqref{equ:maha-distance} is a measure of the distance between a data point and the sample mean, thus, we can remove points that are far from $\Vec{\mu}$ \citep{feigelson2012}.

We use a robust estimator for $\Vec{\mu}$ and $\Vec{S}$ to reduce the effect of the outliers on these values. 
For this, we implemented the Minimum Covariance Determinant (MCD) estimator developed by \citet{rousseeuw1999}.
This method uses a sub-sample of the dataset that minimizes the determinant of the covariance matrix. 
Then, based on the $\chi^2$ distribution and assuming a multivariate normal distribution we can define an outlier threshold, in our case at $c = \sqrt{\chi^2_{p=5}}$ for the $95\%$ confidence interval with $p$ degrees of freedom: $5$ for the astrometric space \verb+5F+. 
Once the threshold is defined, those data points with extreme values (i.e., $d_{M}(\Vec{r_i}) > c$) are removed from the sample. 
Figures \ref{fig:cluster_outliers} and \ref{fig:roslund_6_maha} shows the distribution of the Mahalanobis distances for $486$ samples in the Roslund 6 open cluster selected by the clustering algorithm. 
By following the procedure described above, an outlier threshold of $c=3.33$ was found using the $\chi^2$ distribution.
Those points in Fig. \ref{fig:roslund_6_maha} with values above the cutoff represented by the horizontal dashed blue line are considered outliers. 
We found $57$ stars identified as spurious data in this cluster.
This procedure is applied only on the astrometric space \verb+5F+ with equatorial coordinates, proper motions and parallax corrected by the zero-point bias. 
We opted not to include photometry (\verb|7F|) because white dwarf and red giant stars may be classified as outliers due to their clear separation from the main sequence.


\renewcommand{\arraystretch}{1.2}
\begin{table*}
\caption{Cluster parameters for selected open clusters.}
\label{tab:members}
\centering
\begin{tabular}{@{\extracolsep{4pt}}ccccccccccc@{}}\hline\hline
\multirow{2}{*}{Cluster} & \multirow{2}{*}{$N_{\text{stars}}$} & $\alpha$ & $\delta$ & $\mu_{\alpha *}$ & $\mu_{\delta}$ & $\varpi_P$ & \multirow{2}{*}{log(age)} & $Z$ & $A_{V}$ & $m-M$ \\ \cline{3-4} \cline{5-6} \cline{9-9} \cline{10-11}
&  & \multicolumn{2}{c}{(deg)} & \multicolumn{2}{c}{(mas yr$^{-1}$)} & (mas) &  & (dex) & \multicolumn{2}{c}{(mag)} \\\hline
\multicolumn{11}{c}{$\cdots$} \\
BH 99 & $544$ & $159.49$ & $-59.15$ & $-14.46$ & $1.00$ & $2.207$ & $7.66$ & $0.0672$ & $0.34$ & $8.35$ \\
Collinder 463 & $644$ & $27.06$ & $71.74$ & $-1.74$ & $-0.37$ & $1.123$ & $8.51$ & $0.1943$ & $0.79$ & $9.65$ \\
IC 1396 & $577$ & $324.80$ & $57.53$ & $-2.26$ & $-4.58$ & $1.040$ & $6.98$ & $-0.0225$ & $1.59$ & $9.79$ \\
Melotte 22 & $1130$ & $56.61$ & $24.10$ & $19.92$ & $-45.40$ & $7.322$ & $8.01$ & $0.0287$ & $0.17$ & $5.55$ \\
NGC 1342 & $455$ & $52.91$ & $37.39$ & $0.41$ & $-1.66$ & $1.485$ & $9.05$ & $-0.1134$ & $0.71$ & $9.18$ \\
\multicolumn{11}{c}{$\cdots$} \\\hline
\end{tabular}
\tablefoot{Equatorial coordinates ($\alpha$, $\delta$), proper motions ($\mu_{\alpha *}$, $\mu_{\delta}$) and corrected parallax ($\varpi_P$) are computed as the median values in each cluster sample. The cluster parameters log(age), $Z$, $A_V$ and $m-M$ are estimated through the BASE-9 suite software. The full version of this table will be available at the CDS.}
\end{table*}


\section{Results}
\label{sec:results}

\begin{figure}[t]
    \centering
    \includegraphics[scale=0.77]{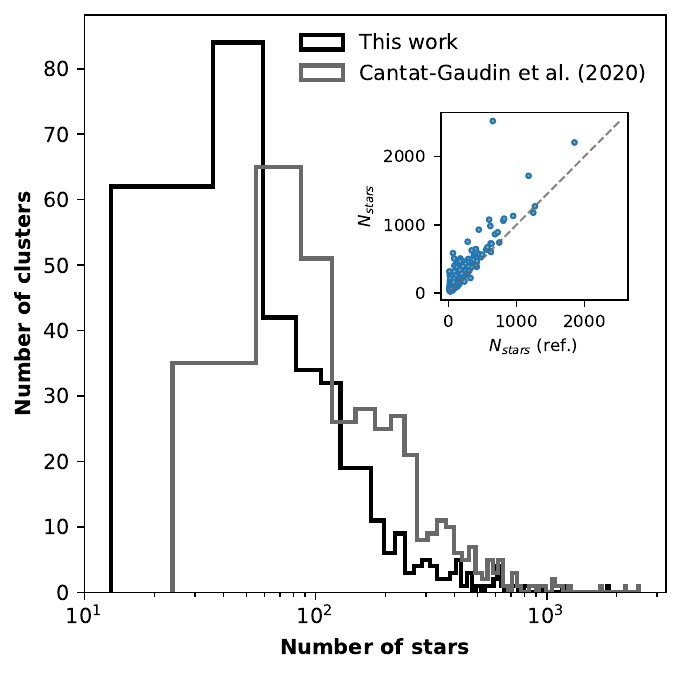}
    \caption{Histogram of the number of stars ($N_{\text{stars}})$ found in this work and those reported in \citet{cantat-gaudin2020}. The insert plot shows the comparison in number of stars between this work ($y$-axis) and \citet{cantat-gaudin2020} ($x$-axis). The dashed line in the insert plot represents the identity.}
    \label{fig:comparison_n_stars}
\end{figure}

In this study we aim to obtain reliable star members for $370$ nearby open clusters, mainly on the Galactic disk, using the {\it Gaia} DR3 catalogue.
We employ the \texttt{HDBSCAN} algorithm applied on two sets: astrometry with five features and a combination of astrometry and photometry with seven features (see Sect. \ref{sec:membership_determination}). 
After clustering application, we cleaned up the samples implementing an outlier removal process with the Mahalanobis metric.
The final stellar members are determined by merging results from \verb+5F+ and \verb+7F+. 
Figure \ref{fig:comparison_n_stars} shows the comparison between the number of stars ($N_{\text{stars}}$) found in this work and those found in the reference sample by \citet{cantat-gaudin2020}.
Table \ref{tab:members} also shows the number of stars and parameters for some clusters found in this work \footnote{The tables with cluster parameters and their stellar members will be made available as online material at the CDS.}
The age, metallicity and extinction were estimated by fitting the PARSEC tracks \citep{bressan2012,marigo2013} to the CMDs through the BASE-9 software.


\subsection{Membership results}
\label{sec:membership_results}

The star membership was performed on {\it Gaia} DR3 using the \texttt{HDBSCAN} algorithm through the \verb+scikit-learn+ Python package \citep{scikitlearn}. 
To select the open clusters and avoid hyperparameters issues that can arise due to the different size populations, we divide the whole sample into five sub-samples as explained in Sect. \ref{sec:data}.
In order to improve the \texttt{HDBSCAN} efficiency, \citet{hunt2023} chose \verb+min_cluster_size+ $\in \{ 10, 20, 40, 80 \}$, which are reasonable sizes for a star cluster. 
Following their methodology, we found that due to the wide population sizes, the same \verb+min_cluster_size+ does not work for all sub-samples.
For that matter, the algorithm was unable to detect open clusters, particularly those with a reduced number of stars ($S_3$, $S_4$, $S_5$). 
Therefore, we chose a different \verb+min_cluster_size+: for $S_1$, $S_2$, $S_3$, $S_4$ and $S_5$ we select $60$, $50$, $40$, $30$ and $20$, respectively.
Moreover, we use \verb+min_samples+ $=5$ instead of $10$ chose by \citet{hunt2023} for all regions.
This further improved the efficiency of recovering the clusters when \texttt{HDBSCAN} was running.
Before clustering implementation, we standardize the sub-samples data through the \verb+RobustScaler+ on the \verb+scikit-learn+ Python package \citep{scikitlearn} scaling features using statistics that are robust to outliers.
Then, the algorithm was applied using the Mahalanobis metric given by Eq. \eqref{equ:maha-metric} as the pairwise metric. 
This increased considerably the computational time due to covariance matrix calculations. 
We ran the entire routines using GPU cores in the supercomputer {\it Hypathia} at Universidad de los Andes, Colombia.
All tasks lasted about eight days per sub-sample.

After the clustering has been applied, we recover the stellar members merging results from \verb+5F+ and \verb+7F+ implementations. 
Then, we remove outliers for the whole sub-samples using once again the Mahalanobis distance given by Eq. \eqref{equ:maha-distance}.
For instance, Fig. \ref{fig:cluster_outliers} shows the Mahalanobis distance distribution in the vector point diagram, equatorial coordinates and the CMD for the Roslund 6 open cluster. 
Following the methodology described in Sect. \ref{sec:outlier_removal} we set an outlier threshold of $3.33$ (dashed horizontal line in Fig. \ref{fig:roslund_6_maha}), then, all data with $d_{M}(\Vec{r_i})$ greater than the threshold are classified as outliers. 
From the Roslund 6 members found by \texttt{HDBSCAN} we obtain a clean list of $429$ stars considering a $95 \%$ confidence interval, which classifies $57$ stars as spurious data. 

It can be noted from Fig. \ref{fig:cluster_outliers} that, expectedly, outliers are located at the outskirts of the cluster and/or at the faint end of the main sequence, where more contamination occurs, about $70\%$ of outliers found have $G > 16$.  Nonetheless, on average for all clusters, about $86\%$ of all initial \texttt{HDBSCAN}-selected members with G>16 are kept as members after the outlier cleaning step, the lowest percentage being $66\%$ and the largest one $100\%$ which occurs for 11 clusters. As an example, for cluster Roslund 6 depicted in Fig. \ref{fig:cluster_outliers}, only 15\% of the faint stars were rejected. This indicates that the methodology applied can properly select bona fide faint cluster members, fulfilling the main goal of this investigation.


\subsection{Approaching the faintest limit}
\label{sec:approaching_the_faintest_limit}

The recent catalogues of open clusters using {\it Gaia} provide valuable information on the census of stars and properties of these Galactic objects.
The efforts are focused on extending the number of stellar members and detecting new clusters based on the astrometric information provided by the catalogues. 
It can be seen how the number of clusters has been growing due to new data collected, from a few thousands in the last decade \citep{karchenko2013} to more than seven thousand in one of the latest catalogues made by \citet{hunt2023}. 
This would not have been possible without {\it Gaia}, its content and its precise parallaxes have changed our view of the Milky Way.

However, no catalogue is completely exempt from systematic errors and dealing with these issues may be tedious. 
Many ways have been used to remove spurious data in the {\it Gaia} catalogues, quality cuts on statistics based on data being the most commonly implemented. 
For instance, \citet{cantat-gaudin2020} opted to choose sources up to $G = 18$ in {\it Gaia} DR2, which lowers the number of stars in the faintest region of the CMD. 
On the other hand, \citet{rybizki2022} has trained neural networks on a diverse set of features for stars with very high signal-to-noise-ratio but negative parallax, i.e., \verb|parallax_over_error < -4.5|, in {\it Gaia} EDR3, which amount to a non-negligible population \citep{luri2018}. 
In addition, to improve the open cluster census, \citet{hunt2023} selected data by cutting on the re-normalised unit weight error (\texttt{ruwe}) and with a quality value of at least $0.5$ in the statistic computed by \citet{rybizki2022}, before clustering implementation.
These cuts on {\it Gaia} DR3 data considerably increased the number of star members in known open clusters and allowed the detection of new ones. 
However, at the faint end in the CMD, parallax uncertainties ($\sigma_\varpi$) increase and if not treated with caution they may distort the spatial distributions \citep{smith1996,bailer2015,luri2018}.
\citet{rybizki2022} also conclude that to obtain good astrometric samples, data can be filtered out by \verb|ruwe| and \verb|astrometric_excess_noise|, which in addition to the \verb|parallax_over_error| may remove spurious sources as done in this study. 
Nevertheless, the selection function is far from trivial and distance estimates inverting the parallax for sources with high parallax uncertainties can bias the results. 
In conclusion, dealing with these issues is far from straightforward and further research is required.

\begin{figure}[t]
    \centering
    \includegraphics[scale=0.7]{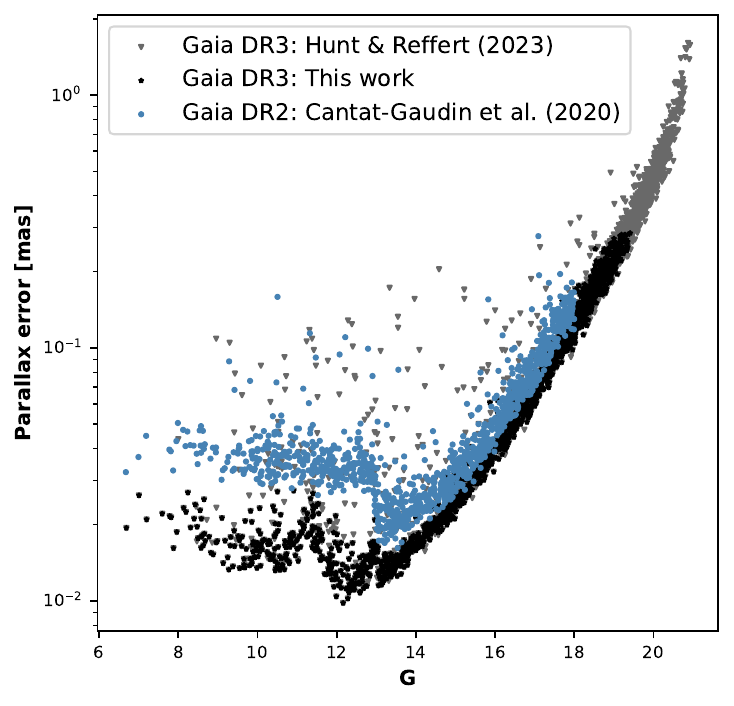}
    \caption{$G$ magnitude vs parallax error ($\sigma_{\varpi}$) for the Stock 2 open cluster. There are $1178$, $1718$, and $3011$ stars classified by \citet{cantat-gaudin2020} (blue points), this work (black points), and \citet{hunt2023} (gray points). The $y$-axis is in logarithmic scale.}
    \label{fig:Stock_2_G_vs_parallax_error}
\end{figure}

In this work, we have aimed to reach a proper balance between extending the faint limit of cluster members while keeping contamination at reduced levels, providing valuable members samples for further meaningful studies of these clusters. Our first step to reduce spurious data is to restrict the {\it Gaia} DR3 data by $\sigma_\varpi$, re-normalised unit weight error and astrometric excess noise (see Sect. \ref{sec:data}).
As mentioned by \citet{lindegren2021}, there are random and systematic errors in the Gaia catalogue, particularly in parallaxes, sources at both bright and faint limits of {\it Gaia} have increasing uncertainties that in case of $\varpi$ will produce poorer estimates of the real parallax $\varpi_{true}$ \citep{luri2018}.

Figure \ref{fig:Stock_2_G_vs_parallax_error} shows the star members for the Stock 2 cluster classified by \citet{cantat-gaudin2020} in blue, this work in black and \citet{hunt2023} in grey. Parallax improvement between Gaia DR2 and DR3 is most remarkable between $G \sim 6$ and $G \sim 15$ but beyond $G \sim 15$ towards the faintest limit, parallax uncertainties still increase considerably. Yet, we have been able to go deeper without introducing significant noise. For example, for the Stock 2 cluster, we found $1374$ stars with $G > 18$ for which the average $\sigma_\varpi$ is $\sim$$0.16$ mas, while the \citet{hunt2023} sample reaches $\sigma_{\varpi}$ up to $\sim$$1.6$ mas, an order of magnitude larger errors. Such high uncertainties may affect the real spatial distribution in the clusters, for example, at a distance of $379$ pc reported for Stock 2 by \cite{cantat-gaudin2020}, a $\sigma_{\varpi} \sim 0.5$ mas would significantly distort its real distance range to 318 $-$ 414 pc, affecting estimates on tidal radius or mass segregation process. On the other hand, a $\sim$$0.16$ mas error corresponds to a difference of $\pm$$25$ pc, below the largest accepted radius for open clusters \citep{portegies2010}.


In addition to this, the spatial distributions for many open clusters with star members from the {\it Gaia} catalogue have line-of-sight stretching. This effect persists even after correcting for bias parallax and estimating the distance using, e.g., the \citet{bailer2015} method. 
This could lead to overestimation of tail-like structures that may be an effect produced by the dynamic evolution of the clusters or merely the product of an optical effect in the parallax measurement.
This problem deserves further investigation for clusters with this effect in their spatial distributions.


\subsection{Inferring the cluster parameters}
\label{sec:stellar_parameter_inference}

The age and metallicity are the cornerstone of stellar evolution, they allow us to know about the chemical composition of the stars in the Milky Way. 
These parameters provide insights into the evolutionary state of stellar systems and also supply an overview of the matter distribution across the Galaxy, which is crucial to understand processes like mass segregation, radial migration, and dynamical heating \citep{mackereth2019}.
Estimating such parameters is a non-trivial task, and the common ways to compute them is by, e.g., lithium depletion boundary and the classical isochrone fitting \citep{dinnbier2022}. 
Furthermore, artificial neural networks have gained popularity and have been successfully implemented in {\it Gaia} data to estimate those parameters bringing us another way to tackle this problem \citep{kounkel2019,cantat-gaudin2020}.

\begin{figure}[t]
    \centering
    \includegraphics[scale=0.38]{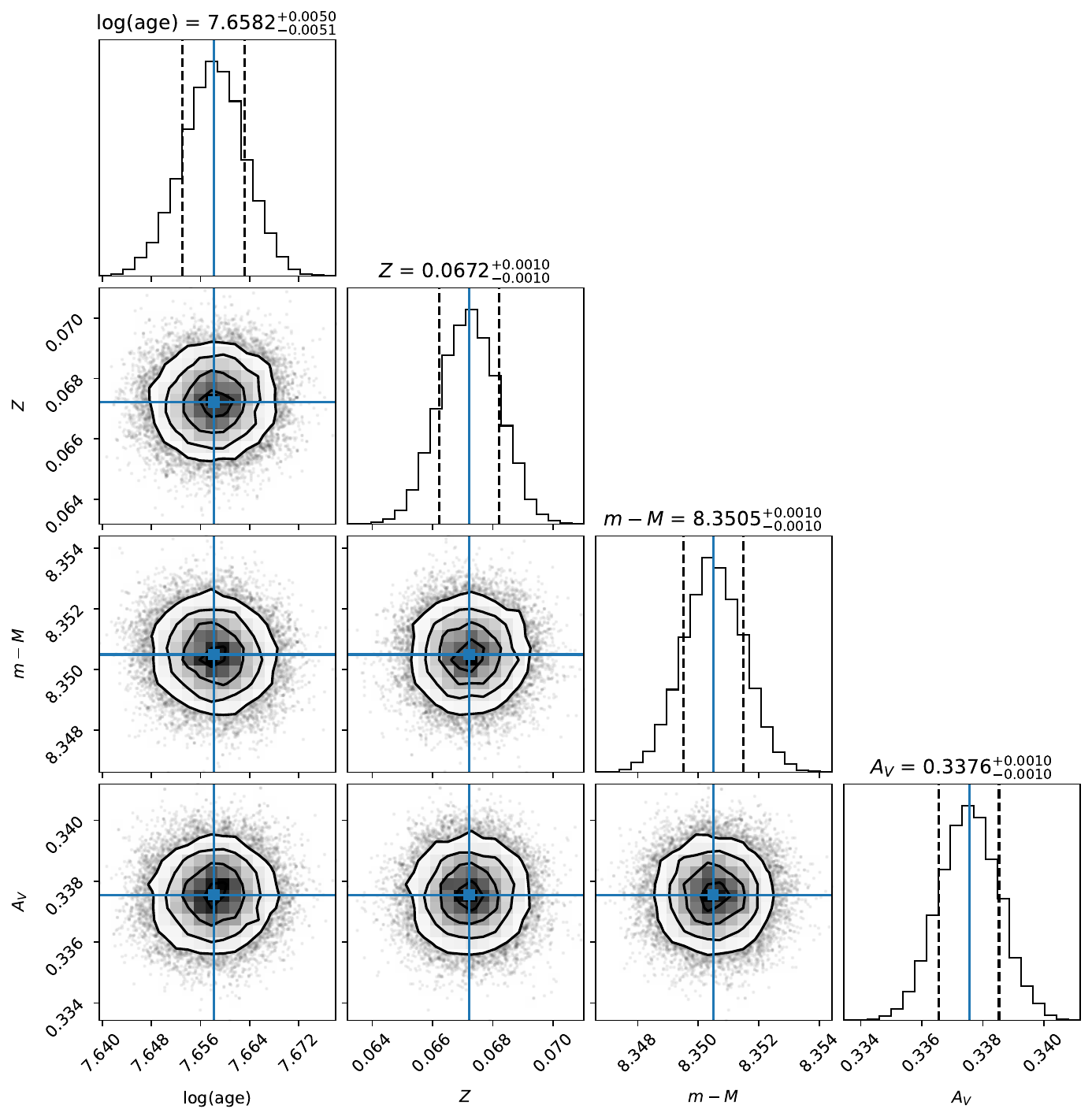}
    \caption{Corner plot showing the marginal posterior probability distribution for the BH 99 open cluster computed using BASE-9 with {\it Gaia} photometry. The blue solid vertical line indicates the estimated parameter, which corresponds to the $50_{\text{th}}$ percentile.}
    \label{fig:corner}
\end{figure}

\begin{figure}[t]
    \centering
    \includegraphics[scale=0.73]{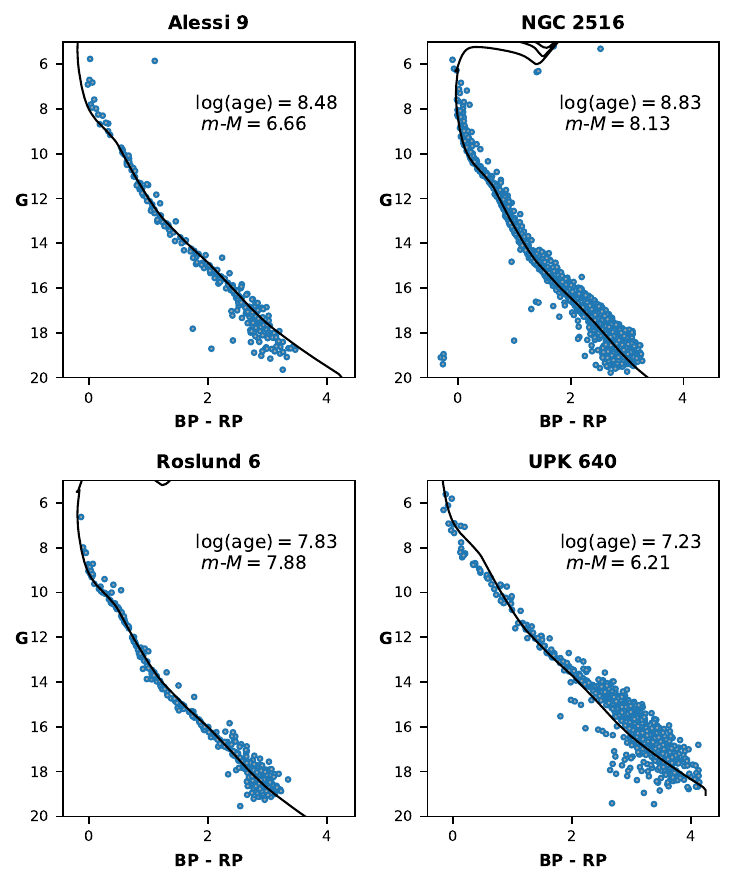}
    \caption{CMDs for Alessi 9, NGC 2516, Roslund 6 and UPK 640 open clusters. The solid black curve is the PARSEC isochrone with parameters estimated through BASE-9.}
    \label{fig:cmd_isochrones}
\end{figure}

To infer age, metallicity, extinction, and distance modulus for each cluster, we opted to fit the PARSEC isochrones \citep{bressan2012,marigo2013} to the CMDs through the Bayesian Analysis for Stellar Evolution with nine variables BASE-9 software \citep{vonhippel2006,robinson2016}. 
This suite software obtains the best \textbf{cluster} parameters by robust statistical principles that compute random walks to sample the marginal posterior probability function. 
BASE-9 requires photometry ($G$, $G_{BP}$, $G_{RP}$) and their uncertainties as inputs to start sampling. 
Since Gaia does not provide photometry errors we use parallax error $\sigma_\varpi$ as a proxy, as done by \citet{kounkel2019}.
In other words, these errors are a measure of relative uncertainties of the data used.
The binaries option in BASE-9 is turned off, therefore all stars are treated as single. 
We chose the initial guess for age, extinction, and distance modulus from \citet{cantat-gaudin2020}.
For metallicity we use the values reported in \citet{netopil2016}, \citet{dias2021} and \citet{fu2022} for a total of $338$ open clusters, the remaining ones have an initial value of $Z=0.01$ dex, which is a reasonable value for open clusters in the solar neighborhood \citep{bossini2019}.
BASE-9 was executed on blocks of $40$ clusters in parallel with a total of one hundred thousand iterations.
After sampling, we obtain a flat chain with a length of ten thousand iterations, thus the estimated cluster parameters are the $50_{th}$ percentiles or the second quartiles $Q_2$.
Figure \ref{fig:corner} shows the corner plot of the marginal posterior probability function sampling for the BH 99 cluster with the upper and lower uncertainties in each parameter as the $16_{th}$ and $84_{th}$ percentiles. 
Some isochrones for Alessi 9, NGC 2516, Roslund 6 and UPK 640 open clusters with parameters estimated through BASE-9 are depicted in Fig. \ref{fig:cmd_isochrones}.

\begin{figure*}[t]
    \centering
    \includegraphics[scale=0.8]{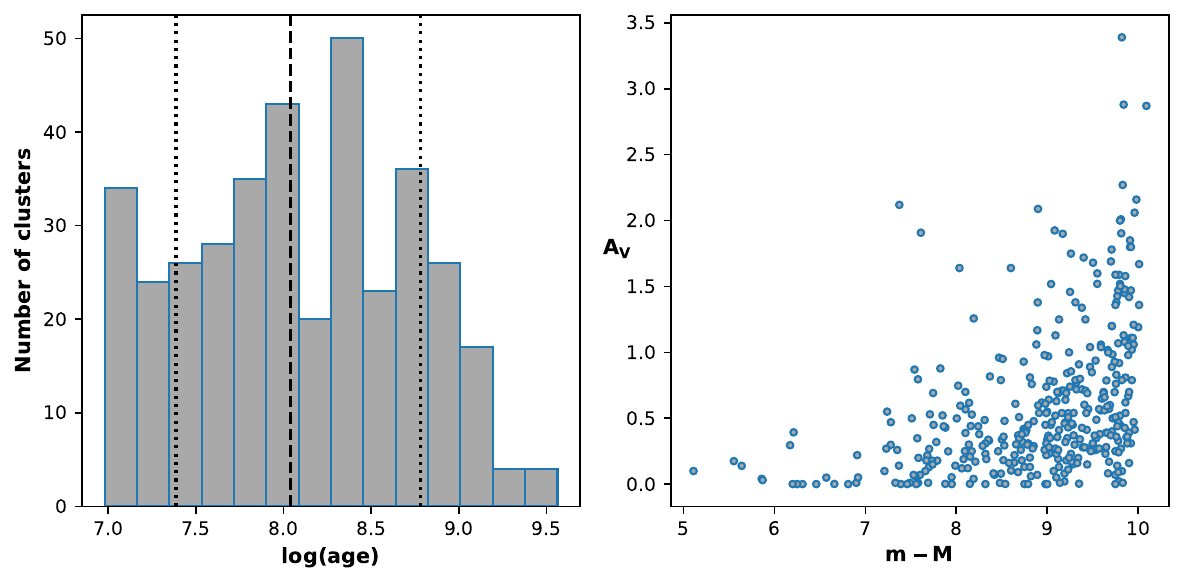}
    \caption{{\it Left panel}: Age distribution of the studied open clusters, the dashed vertical line represents the median value (log(age)$=8.03$), while the dotted lines are the $16_{\text{th}}$ (log(age)$=7.38$) and $84_{\text{th}}$ (log(age)$=8.78$) percentiles, respectively. {\it Right panel}: Extinction versus distance modulus estimated in this work.}
    \label{fig:stellar_parameters_results}
\end{figure*}

\begin{figure*}[t]
    \centering
    \includegraphics[scale=0.9]{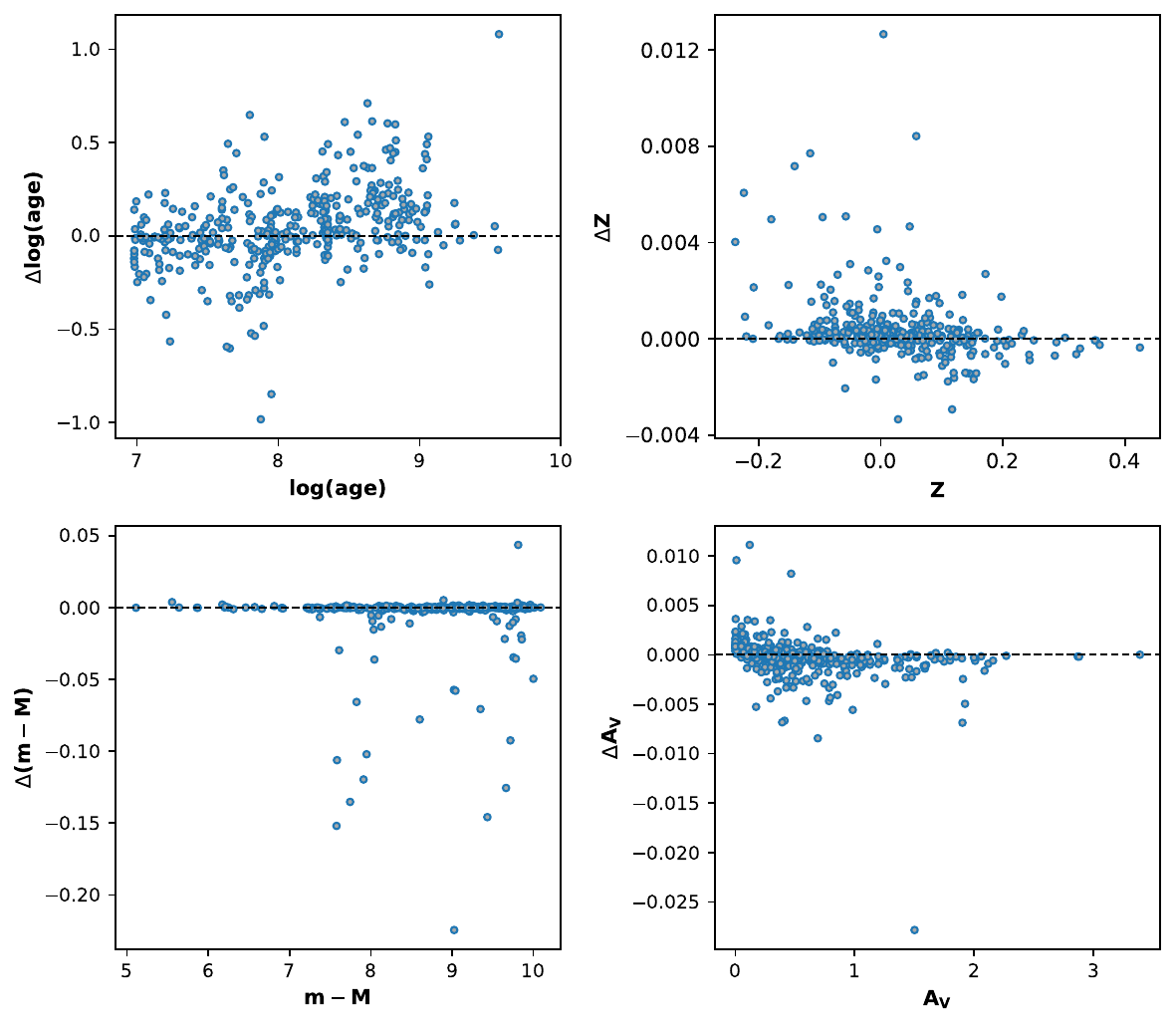}
    \caption{Comparison of cluster parameters derived in this work. 
    The log(age) ({\it top left panel}), distance modulus ({\it bottom left panel}) and extinction ({\it bottom right panel}) are compared against \citet{cantat-gaudin2020} for $370$ clusters.
    The metallicity ({\it top right panel}) is compared against \citet{dias2021} for $324$ clusters in common.
    $\Delta$log(age) $=$ log(age)$_{\text{this work}}$ - log(age)$_{\text{literature}}$ versus log(age)$_{\text{this work}}$. 
    $\Delta$Z $=$ Z$_{\text{this work}}$ - Z$_{\text{literature}}$ versus Z$_{\text{this work}}$. 
    $\Delta(m-M)$ $=$ $(m-M)_{\text{this work}}$ - $(m-M)_{\text{literature}}$ versus $(m-M)_{\text{this work}}$.
    $\Delta A_V$ $=$ $A_V$$_{\text{this work}}$ - $A_V$$_{\text{literature}}$ versus $A_V$$_{\text{this work}}$.}
    \label{fig:stellar_comparison}
\end{figure*}

The left panel of Fig. \ref{fig:stellar_parameters_results} presents the age distribution of the studied open clusters. 
The log(age) cover a range from $6.97$ with Collinder 69 as the youngest, to $9.56$ with UBC 21 as the oldest cluster in the sample. 
Among the $370$ clusters, $250$ open clusters were found between the $16_{\text{th}}$ and $84_{\text{th}}$ percentiles with a median value of log(age)$=8.03$, similar to the median of log(age)$=8.2$ in \citet{bossini2019}.
The right panel of Fig. \ref{fig:stellar_parameters_results} shows the extinction versus distance modulus, which shows that most of the clusters have low extinction with values less than $1.0$ mag.
The UPK 201 cluster presents the highest extinction ($A_V=3.39$) in our sample.
Figure \ref{fig:stellar_comparison} shows the comparison for age ({\it left top panel}), distance modulus ({\it left bottom panel}) and extinction ({\it right bottom panel}) estimated in this work against \citet{cantat-gaudin2020} (for all clusters), and for metallicity  ({\it right top panel}) against \citet{dias2021} (for $324$ open clusters in common).
\citet{cantat-gaudin2020} values were obtained by an artificial neural network trained using the PARSEC isochrones. \citet{dias2021} metallicities were obtained using the {\it Gaia} DR2 photometry and an isochrone fitting code also with the PARSEC tracks.

As explained before, \citet{cantat-gaudin2020} opted to perform a cutoff in $G \sim 18$ in order to avoid issues previously stated (see Sect. \ref{sec:approaching_the_faintest_limit}) in the fainter region of the CMDs, while we include stellar members with magnitudes up to $G \sim 19.8$ in most of the clusters using quality cuts discussed in Sect. \ref{sec:data}.
We found slight differences in ages compared to \citet{cantat-gaudin2020} that may have different causes, such as different methods to estimate age and our work including fainter members.
Nevertheless, the discrepancies are limited to a few values and only $18$ clusters have $|\Delta\text{log(age)}| > 0.5$.
Particularly, the three open clusters with the highest discrepancies in $\Delta$log(age) are UBC 21, UPK 18 and UPK 542. Nonetheless, these clusters agree in the other parameters, suggesting that age estimates are sensitive to large scatter in the CMD.
In addition, most of the clusters have similarities in extinction with \citet{cantat-gaudin2020} and only $10$ of them have $|\Delta A_V| > 0.005$.
In case of distance modulus, the discrepancies are small and only $14$ open clusters have $|\Delta(m-M)| > 0.05$, the vast majority of them with distance estimates 10\% or less closer than those of \citet{cantat-gaudin2020}.

We note that only $27$ among the $324$ clusters have $|\Delta\text{Z}| > 0.002$ dex with a median value of $0.0001$ dex, which indicates similarities in the metallicities derived in this work with \citet{dias2021}.
Further research is required to quantify the impact of the inclusion of faint and low-mass stars on age and metallicity estimates through the standard isochrone fitting process.
This will be discussed in the second series of this paper (Alfonso et al. in prep.).



\begin{figure*}[t]
    \centering
    \includegraphics[scale=0.75]{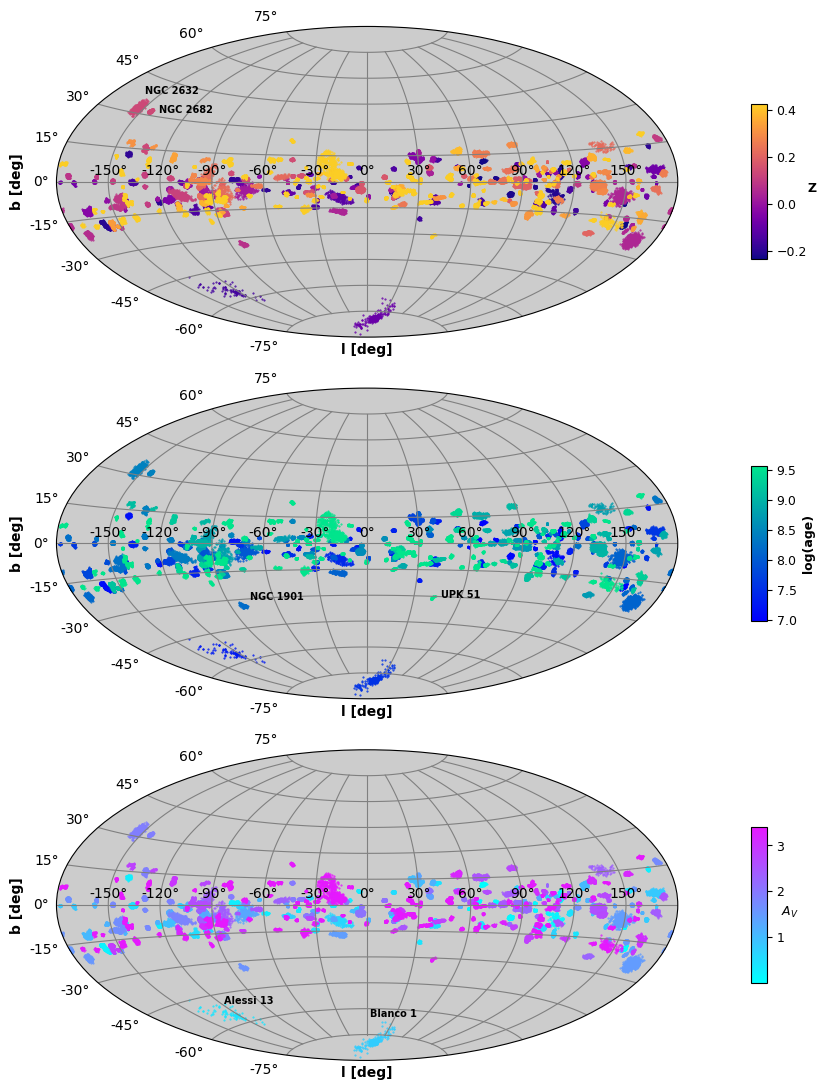}
    \caption{Celestial maps in Galactic coordinates for the $370$ open clusters with a total of $87708$ stars. The color bar indicates the metallicity ({\it top panel}), log(age) ({\it middle panel}) and extinction ({\it bottom panel}) distribution estimated in this work. The open clusters NGC 2632, NGC 2682, UPK 51, NGC 1901, Alessi 13 and Blanco 1 have the highest and lowest Galactic latitude.}
    \label{fig:stellar_distribution}
\end{figure*}


\section{Discussion}
\label{sec:discussion}

We have updated the cluster parameters with the {\it Gaia} DR3 catalogue for $370$ nearby open clusters in which we obtain reliable stellar members using the \texttt{HDBSCAN} clustering algorithm with the Mahalanobis distance as the pairwise metric.
Among the cluster sample we cover a distance range from $\sim 104$ pc with Alessi 13 depicted in the bottom panel of Fig. \ref{fig:stellar_distribution} as the closest, to $\sim 998$ pc with NGC 7762 as the farthest.
We show in Fig. \ref{fig:data_samples} that these clusters are mainly in the Galactic plane, which in fact, the colors for the five sub-samples highlights the wide variety of cluster size populations, as mentioned in Sect. \ref{sec:data}.
We also cover metallicities ranging from $-0.23$ dex to $0.42$ dex with UBC 31 as the most metal-rich open cluster.
Moreover, the ages calculated in this work are in the range of $6.98$ to $9.56$ with Collinder 69 and UBC 21 as the youngest and oldest open clusters, respectively. 
The latter with the highest age increment of $+1.08$ log(age) compared to \citet{cantat-gaudin2020}.

The cluster parameters, shown in Table \ref{tab:members}, were estimated with the PARSEC isochrones through the BASE-9 implementation. 
Figure \ref{fig:stellar_distribution} shows the obtained metallicity, age, and extinction of the open clusters, color-coded in a sky map.
A total of $53890$ of our $87708$ stellar members are in the catalogue provided by \citet{cantat-gaudin2020}, representing an increment of about $\sim 40\%$.
It is worth mentioning that there are $1738$ repeated stars in the full sample listed as members of two or three clusters.
This may be caused by very close binary open clusters that the algorithm overlap as one and requires further analysis case by case to reach a conclusion.

On the other hand, the current census in \cite{hunt2023} has about seven thousand clusters with almost one thousand five hundred within $1$ kpc. 
We opted to focus on open clusters not including comoving groups and streams as Theias \citep{kounkel2019}, their wide variety of stellar content not sharing a common origin \citep{zucker2022} may be challenging for cluster parameter estimation. 
A total of $56$ and $46$ open clusters have now parameters not reported previously by \cite{hunt2023} and \cite{dias2021}, respectively.

In our sample, we found four pairs of truly binary open clusters reported by \citet{delafuente2009} and \citet{song2022}. These pairs have distances about $30$ pc between the components and also similar ages and metallicities.
They are: Collinder 135 - UBC 7, Alessi 43 - Collinder 197, NGC 6716 - Collinder 394 and ASCC 16 - ASCC 21.
In addition, we also found six system pairs with distances less than about $20$ pc: ASCC 123 - Stock 12, BDSB91 - vdBergh 80, Collinder 394 - NGC 6716, Gulliver 6 - UBC 17b, RSG 7 - RSG 8 and UBC 392 - UPK 194.
These objects deserve further research to confirm or discard a common origin or physical association \citep{delafuente2009}.

We note that including photometry to \texttt{HDBSCAN} allows to obtain faint stars that were not detected using only astrometry, with parallax uncertainties up to $\sigma_{\varpi} = 0.16$ mas for the open clusters studied in this work. 
These faint stars are crucial to understand dynamical properties of these clusters, such as the mass segregation effect, which may be due to dynamical evolution from a non-mass segregated cluster or primordial as a product of star formation process \citep{richard2009}. 
Nevertheless, including faint stars with the {\it Gaia} data is a cautious task due to the astrometric uncertainties: the fainter in the CMD, the larger the $\sigma_\varpi$ because of the photon limit in the {\it Gaia} CCD detector (see Fig. \ref{fig:Stock_2_G_vs_parallax_error}). 
At the faint regime, the main sequence has high dispersion, which can make it difficult to estimate cluster parameters.
Further investigation is require to quantify how well or poorly age and metallicity are estimated through traditional isochrone fitting including these stars beyond $G \geq 20$. 
In addition, using quality cuts such as those adopted in this work (see Sect. \ref{sec:data}) can deal with this problem \citep{lindegren2021,rybizki2022}, which reduces the number of spurious data that may affect cluster parameter estimates and clustering results.

This study is mainly focused on obtaining star members for open clusters near the Galactic plane within $1$ kpc with the novel {\it Gaia} DR3 catalogue using the unsupervised machine learning algorithm \texttt{HDBSCAN} applied on astrometry and photometry. 
The algorithm was applied taken into account the correlations among the different kind of informations provided in the catalogue through the Mahalanobis metric and also used the latter to remove outliers from the cluster samples.
We found stars up to $G \sim 19.8$ that were not included in the reference \citet{cantat-gaudin2020} sample with {\it Gaia} DR2. 
The exquisite information in {\it Gaia} DR3 allows us to compute basic kinematic parameters and also update age, extinction, distance modulus, and determine metallicity which was not reported in the reference sample. 
Additionally, spectroscopic surveys such as LAMOST \citep{lamost2012} and WEAVE \citep{weave2012} with {\it Gaia} will provide an extended and full characterization of the dynamical and stellar process in these open clusters, which for the lack of radial velocities in {\it Gaia} will bring full details about the kinematic of these clusters in their velocity spaces.


\section{Summary and conclusions}
\label{sec:conclusions}
In this work, we have obtained reliable star members for $370$ nearby open clusters using the novel {\it Gaia} DR3 and the \texttt{HDBSCAN} algorithm applied in two approaches. 
First, we performed the algorithm on the common astrometric space \verb|5F| with equatorial coordinates, proper motions, and the parallax corrected by the zero-point bias. 
Then, we apply once again \texttt{HDBSCAN} by adding two color indexes \verb|7F|.
We note that by including photometry to clustering, the number of faint stars with $G>17$ increases about $10\%$ compared to clustering based solely on astrometry, which allows us to obtain additional stars in the faintest region of the CMD.
The Mahalanobis metric was used as the pairwise metric to compute distances for clustering to take into account any level of correlations between astrometric and photometric data.
This metric was also used to remove outliers from the membership lists, we compute the covariance matrix through the Minimum Covariance Determinant estimator to avoid influences of extreme values.
Then, based on the $\chi^2$ distribution and considering a $95\%$ of confidence interval with $5$ degrees of freedom for the astrometric space \verb|5F|, we remove spurious data from the membership lists in each cluster.

We used the {\it Gaia} DR3 photometry with the stellar members classified by the clustering algorithm to estimate age, metallicity, extinction and distance modulus fitting the PARSEC isochrones through the BASE-9 suite software.
We observe that to establish the metallicity distribution in open clusters, it is required to extend the membership list to clusters beyond the solar neighborhood. 
This will bring details about the structure of the Galaxy and also chemical distribution in the thin and thick disk.
In this work, we show the need to carefully select stars in open clusters, which due to astrometric and photometric uncertainties in {\it Gaia}, may affect clustering results, produce an overestimation of the cluster parameters because of scattering at faint magnitudes, and also identify real or fake stretching in the line of sight of the clusters. These issues will be discussed in the second series of this paper (Alfonso et al. in prep.).

The catalogue presented in this paper exploits the {\it Gaia} DR3 data to extent stellar members in nearby clusters without including stars with high parallax uncertainties. 
We found a total of $87708$ stars for the $370$ open clusters studied in this work.
This first paper also provides a new technique to tackle the membership problem in star clusters including astrometry and photometry data.
Our approach relies on dealing with correlations in the data and use robust statistical estimator to avoid any level of affectation in the final membership lists.


\begin{acknowledgements}
The authors thank the anonymous referee for his/her/their very valuable and constructive comments and suggestions that improve the quality of this manuscript.
The authors would like to thank the Vice Presidency of Research \& Creation’s Publication Fund at Universidad de los Andes for its financial support and also the Fondo de Investigaciones de la Facultad de Ciencias de la Universidad de los Andes, Colombia, through its Programa de Investigación c\'{o}digo INV-2023-162-2853.
Jeison Alfonso acknowledges doctoral fellowship support from the Departamento de Física de la Universidad de los Andes, Colombia, in form of {\it Asistencia Graduada Doctoral Docencia}.
This work has made use of data from the European Space Agency (ESA) mission {\it Gaia} (\url{https://www.cosmos.esa.int/gaia}), processed by DPAC, (\url{https://www.cosmos.esa.int/web/gaia/dpac/consortium}). Funding for the DPAC has been provided by national institutions, in particular the institutions participating in the {\it Gaia} Multilateral Agreement.
\end{acknowledgements}


\bibliographystyle{aa}
{\footnotesize\bibliography{bibliography.bib}}






\end{document}